# Magnetic phase diagram of the antiferromagnet $U_2Rh_2Pb$

J. Pospíšil[a,*], M. Míšek[b], M. Diviš[a], M. Dušek[b], F.R. de Boer[a], L. Havela[a], J.Custers[a]

[a] *Charles University, Faculty of Mathematics and Physics, Department of Condensed Matter Physics, Ke Karlovu 5, 121 16 Prague 2, Czech Republic*
[b] *Institute of Physics, Academy of Sciences of Czech Republic, v.v.i, Na Slovance 2, 182 21 Prague 8, Czech Republic*

*Corresponding author: jiri.pospisil@mag.mff.cuni.cz (J. Pospíšil)

**Abstract**

A new U-based compound of the $U_2Rh_2Pb$, a new compound of the $U_2T_2X$ series (T – transition metal, X – Sn, In, Pb), was prepared in a single-crystal form. Its structure was determined as belonging to the tetragonal $Mo_2FeB_2$ structure type with the shortest U-U spacing along the *c*-axis. The crystals were subjected to study of magnetic, specific heat, and electrical resistivity in various magnetic fields. $U_2Rh_2Pb$ undergoes an antiferromagnetic transition at a Néel temperature $T_N$ of 20 K and exhibits an enhanced Sommerfeld coefficient $\gamma \approx 150$ mJ/molK$^2$. In contrast to the two rhodium analogues $U_2Rh_2In$ and $U_2Rh_2Sn$, the easy-magnetization direction is the *c*-axis with rather low value of the critical field $H_c = 4.3$ T of the metamagnetic transition of a spin-flip type. The observed dependences of $T_N$ and $H_c$ on temperature and magnetic field have been used for constructing a magnetic phase diagram. The experimental observations are mostly supported by first-principles calculations.

## 1. Introduction

Compounds of composition $RE(A)_2T_2X$, (RE = rare-earth element, or A = actinide, T = transition metal and X = p-element), crystallizing in the tetragonal $Mo_2FeB_2$ type of structure with space group *P4/mbm* [1], represent a large group, which has been intensively studied due to the extraordinary freedom to combine the various constituting elements while maintaining the structure type. Thus, both the rare-earth and the actinide compounds offer unique opportunity for systematic investigation of the magnetic and electronic phenomena as a function of lattice parameters and chemical composition [2-4].

Uranium compounds, one of which is subject of this paper, display a large diversity of the types of ground state, including itinerant paramagnets ($U_2Co_2In$, $U_2Ru_2Sn$), some with pronounced spin-fluctuation behavior ($U_2Co_2Sn$ [5] and $U_2Ir_2In$ [2]), suspected Kondo behavior ($U_2Rh_2In$ [6]) and long-range antiferromagnetic (AFM) ordering ($U_2Ni_2In$, $U_2Ni_2Sn$, $U_2Rh_2Sn$, $U_2Pd_2In$, $U_2Pd_2Sn$ and $U_2Pt_2Sn$) [2, 7]. Some of the $U_2T_2X$ compounds, like $U_2Pt_2In$ [8] and $U_2Pd_2In$ [4], exhibit heavy-fermion-phenomena at low temperatures. Computational studies of $U_2Pd_2In$ demonstrated that the AF structure and magnetic anisotropy is a result of interplay of geometrical frustration of the lattice and strong spin-orbit coupling with a contribution of Dzyaloshinskii-Moriya interaction [9, 10].

From the large series of $U_2T_2X$ compounds, so far practically only In and Sn compounds have been systematically studied. The only Pb compound is $U_2Pd_2Pb$, which orders antiferromagneticaly at $T_N$ = 65 K and undergoes an order to order transition at $T_1$ = 20 K [11]. However, $U_2Pd_2Pb$ has been prepared only in polycrystalline form so that the physical properties which involve magnetic anisotropy could not been studied.

In the present study we describe synthesis, structure, and properties of another Pb compound, $U_2Rh_2Pb$, successfully prepared in the single-crystalline form. This compound, which undergoes antiferromagnetic ordering, is unique in several aspects. The *c*-axis orientation of U moments occurs only seldom for $U_2T_2X$ compounds. Moreover, $U_2Rh_2Pb$ exhibits the lowest critical magnetic field of the field-induced (metamagnetic) transition (MT). The experimental findings amount into construction of magnetic phase diagram. The experimental observations are compared with results of first-principles calculations.

## 2. Experimental and theoretical methods

$U_2Rh_2Pb$ single crystals have been grown by the self-flux method from high-purity elements U (3N purified by SSE [12]), Rh (4N5) and Pb (5N) in an alumina crucible sealed under vacuum ($10^{-6}$ mbar) in a quartz ampoule. Various ratios of the elements were tested. The best results were obtained with the starting compositions U:Rh:Pb=1:1:15 and 1:1:25. A simple temperature profile was set up. The ampoules were heated up to 1000°C, kept at this temperature for 10 h to let the mixture homogenize properly and consequently cooled down to 700°C in 100 h. After decanting of the residual lead, needle-like single crystals of $U_2Rh_2Pb$ with typical dimensions of 2×0.1×0.1 mm$^3$ were obtained. A secondary binary phase of composition $UPb_3$ was also found in the form of small (1×1×1 mm$^3$) cubes. The chemical composition of samples was verified by a scanning electron microscope Tescan Mira I LMH, equipped with an energy dispersive Bruker AXS X-ray detector. The crystal structure and orientation of the needles were determined by single crystal X-ray diffraction using a Gemini X-ray diffractometer, equipped with a Mo lamp, a graphite monochromator and a Mo-enhanced collimator producing Mo radiation and a CCD detector Atlas. Absorption correction of the strongly absorbing samples was done by combination of numerical absorption correction based on the crystal shapes and empirical absorption correction based on spherical harmonic functions, using the software of the diffractometer CrysAlis PRO. The crystal structures were resolved by SUPERFLIP [13] and refined by Jana2006 [14]. PPMS9T and PPMS14T equipment was used for the specific-heat ($C_p$) and electrical-resistivity ($\rho$) measurements. Because of the small mass of the needles, many single crystals were arranged on a specific-heat-measurement puck, fixed by Apiezon N grease. In this setting, the basal planes of the tetragonal crystals are randomly arranged parallel to the external magnetic field. The measurement was performed by the relaxation method. One long needle of length 4mm$^3$ was extracted for electrical-resistivity measurement by the a.c. four-probe method in longitudinal arrangement. Both the electrical current and the magnetic field were applied along the $c$-axis. Magnetization measurements were performed in a commercial MPMS7T device on a sample consisting of a number of needles arranged parallel in order to increase the signal.

The magnetic moments, magnetocrystalline-anisotropy energy (MAE), and equilibrium volume were calculated using computational methods based on DFT. It is important to employ a full-potential method since the physical properties of $U_2Rh_2Pb$ are very anisotropic. The computer codes Full Potential Local Orbitals (FPLO) [15] and Full Potential Augmented Plane Waves plus local orbitals (APW+lo) [16] were used to solve the single-particle Kohn-Sham equations. The 5f states were treated as itinerant Bloch states in both methods. The fully relativistic Dirac four-component option was used in all FPLO calculations. For testing purposes, scalar relativistic calculations were conducted for both FPLO and APW+lo resulting in the same spin magnetic moment of the unit cell at experimental equilibrium which confirms the full compatibility of both methods. For other details of DFT calculations see Refs. [17-19]. The calculations with the FPLO code were done with the local-spin-density approximation (LSDA) [20] and the generalized gradient approximation (GGA) [21] whereas, in the case of APW+lo, the two additional GGA methods [22, 23] were used.

## 3. Results and discussion
### 3.1 Crystal structure

A series of needle-like U$_2$Rh$_2$Pb single crystals (Fig. 1) has been separated, on which electron microprobe analysis was performed in order to determine the composition. Pb microdrops remaining after the decanting were locally detected on the surface of the needles.

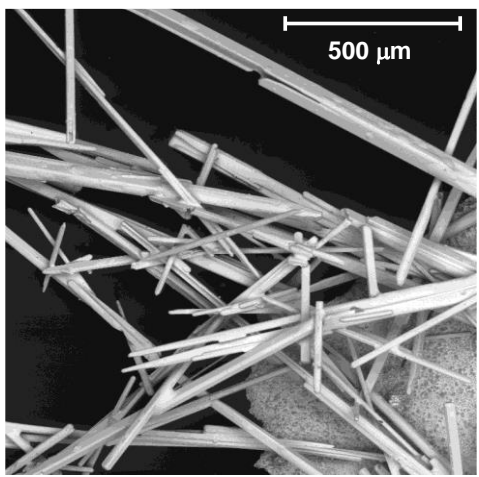

Fig. 1. Morphology of the U$_2$Rh$_2$Pb single crystals in a dense needle nest.

The tetragonal lattice of the Mo$_2$FeB$_2$ structure type with space group *P4/mbm* was confirmed using the single-crystal X-ray diffraction. Crystal structure was evaluated based on recorded total 3055 reflections, 170 recorded independent ones with agreement factors $R[F^2 > 3\sigma(F^2)] = 0.024$, $wR(F^2) = 0.090$, $S = 1.33$. The unit-cell volume of U$_2$Rh$_2$Pb is larger than that of the two other Rh analogues, which can be related to the larger size of the Pb atoms [24] (Table 1). The larger volume is primarily realized by expansion of the tetragonal basal plane, which is partly compensated by a shrinking of the lattice parameter *c*. The shortest uranium-atom distance $d_{U-U}$ is within the U-atoms chain along the *c* axis and is identical to the lattice parameter *c*. This means that the critical distance $d_{U-U}$ in U$_2$Rh$_2$Pb is shorter than in the analogue In and Sn compounds but still within the magnetic limit of the empirical Hill criterion [25]. In the basal plane, each U atom has 1 U neighbor at 3.700 Å and 4 neighbours at 4.01 Å. Comparison of the lattice parameters of all Rh analogues is presented in Table 1. The fractional coordinates are listed in Table 2. Every type of atom has one unique site. Selected interatomic distances are summarized in Table 3.

**Table 1**. Lattice parameters of U$_2$Rh$_2$X with X = In, Sn and Pb.

|   | *a* (Å) | *c* (Å) | *V* (Å$^3$) |
|---|---|---|---|
| U$_2$Rh$_2$In [4] | 7.553 | 3.605 | 205.66 |
| U$_2$Rh$_2$Sn [4] | 7.529 | 3.635 | 206.05 |
| U$_2$Rh$_2$Pb | 7.6478(9) | 3.5899(5) | 209.97(5) |

**Table 2**. Atomic coordinates in the U$_2$Rh$_2$Pb crystal structure (all crystallographic sites are fully occupied).

| Atom | Site | *x* | *y* | *z* |
|---|---|---|---|---|
| U | 4h | 0.32895(8) | 0.82895(8) | 0.5 |
| Rh | 4g | 0.1307(2) | 0.6307(2) | 0 |

|  | Pb | 2a | 0.5 | 0.5 | 0 |

**Table 3**. Selected interatomic distances in the U$_2$Rh$_2$Pb crystal structure. Symmetry codes: (i) $x, y, z-1$; (ii) $x, y, z+1$; (iii) $-x+1, -y+2, z$; (iv) $-y+1, x, z$; (v) $-y+1, x+1, z$; (vi) $y-1, -x+1, z$; (vii) $y, -x+1, z$; (viii) $-x+1/2, y+1/2, -z$; (ix) $-x+1/2, y+1/2, -z+1$; (x) $-y+1, x+1, z+1$; (xi) $y, -x+1, z+1$; (xii) $-x+1, -y+1, z$; (xiii) $-x, -y+1, z$.

| Sites | d (Å) | Sites | d (Å) |
|---|---|---|---|
| U1—U1$^i$ | 3.5899 (10) | U1—Rh1$^{ii}$ | 2.7963 (13) |
| U1—U1$^{ii}$ | 3.5899 (10) | U1—Rh1$^v$ | 2.9398 (14) |
| U1—U1$^{iii}$ | 3.7000 (11) | U1—Rh1$^x$ | 2.9398 (14) |
| U1—U1$^{iv}$ | 4.0100 (12) | U1—Rh1$^{vii}$ | 2.9398 (14) |
| U1—U1$^v$ | 4.0100 (12) | U1—Rh1$^{xi}$ | 2.9398 (14) |
| U1—U1$^{vi}$ | 4.0100 (12) | Pb1—Pb1$^i$ | 3.5899 (10) |
| U1—U1$^{vii}$ | 4.0100 (12) | Pb1—Pb1$^{ii}$ | 3.5899 (10) |
| U1—Pb1 | 3.3559 (7) | Pb1—Rh1 | 2.9960 (17) |
| U1—Pb1$^{ii}$ | 3.3559 (7) | Pb1—Rh1$^{xii}$ | 2.9960 (17) |
| U1—Pb1$^{viii}$ | 3.3559 (7) | Pb1—Rh1$^{iv}$ | 2.9960 (17) |
| U1—Pb1$^{ix}$ | 3.3559 (7) | Pb1—Rh1$^{vii}$ | 2.9960 (17) |
| U1—Rh1 | 2.7963 (13) | Rh1—Rh1$^{xiii}$ | 2.8270 (20) |

### 3.2 Specific heat

In order to obtain sufficient mass for a proper signal for measurement of the specific heat, tens of needles were collected on the specific-heat-measurement puck. The specific heat was measured with the magnetic field applied along the randomly oriented basal planes (field perpendicular to needles) and along the $c$ axis (field along the needles). In the zero-field curve, a λ-type of anomaly, indicating a magnetic phase transition, was detected at 20.5 K (Fig. 2a). Upon increasing field perpendicular to needles, the anomaly shifts to lower temperatures, which is characteristic for AFM order. At 14 T, a small shift of about 1 K was detected (Fig. 2a). An estimate of the critical field where the AFM order would vanish gives over hundred tesla, which is significantly higher than in related U$_2$T$_2$X compounds [2, 26, 27]. This indicates that the basal plane is the hard magnetization direction. Extrapolation of the specific heat plotted as $C_p/T$ to zero temperature provides a Sommerfeld coefficient of electronic specific heat $\gamma = 150 \pm 20$ mJ/mol f.u.K$^2$ (mole is related to the formula unit throughout the paper) (Fig. 2b) which indicates an enhanced density of states at the Fermi level. Both $T_N$ and $\gamma$ are similar to the values reported for the AFM analogue U$_2$Rh$_2$Sn while a much higher $\gamma$ value of 280 mJ/molK$^2$ has been found for the non-magnetic compound U$_2$Rh$_2$In [7].

To determine the magnetic entropy $S_{mag}$ associated with the AF ordering, the low-temperature phonon part $C_{ph}$ has been tentatively separated expressing the specific heat as $C_p/T = \gamma + \beta T^2$ as commonly done for uranium intermetallics [27-30]. However, we could not find any reasonable validity range for this type of dependence. Since, so far, a non-magnetic lead analogue has not yet been discovered, a general polynomial function has been used to estimate the background phonon part $C_{ph}$ below $T_N$ (Fig. 2b). The evaluated value of the magnetic entropy $S_{mag} = 0.3*ln2$ indicates the 5$f$ band magnetism. In reality this value represents a higher estimate. In itinerant magnetic materials, the effective paramagnetic $\gamma$-value tends to be higher than the real $C_p/T$ in the low temperature limit. Integrating the magnetic contribution of $C_p/T$ close to $T = 0$ gives actually a negative contribution, so that $S_{mag}$ can approach values close to zero.

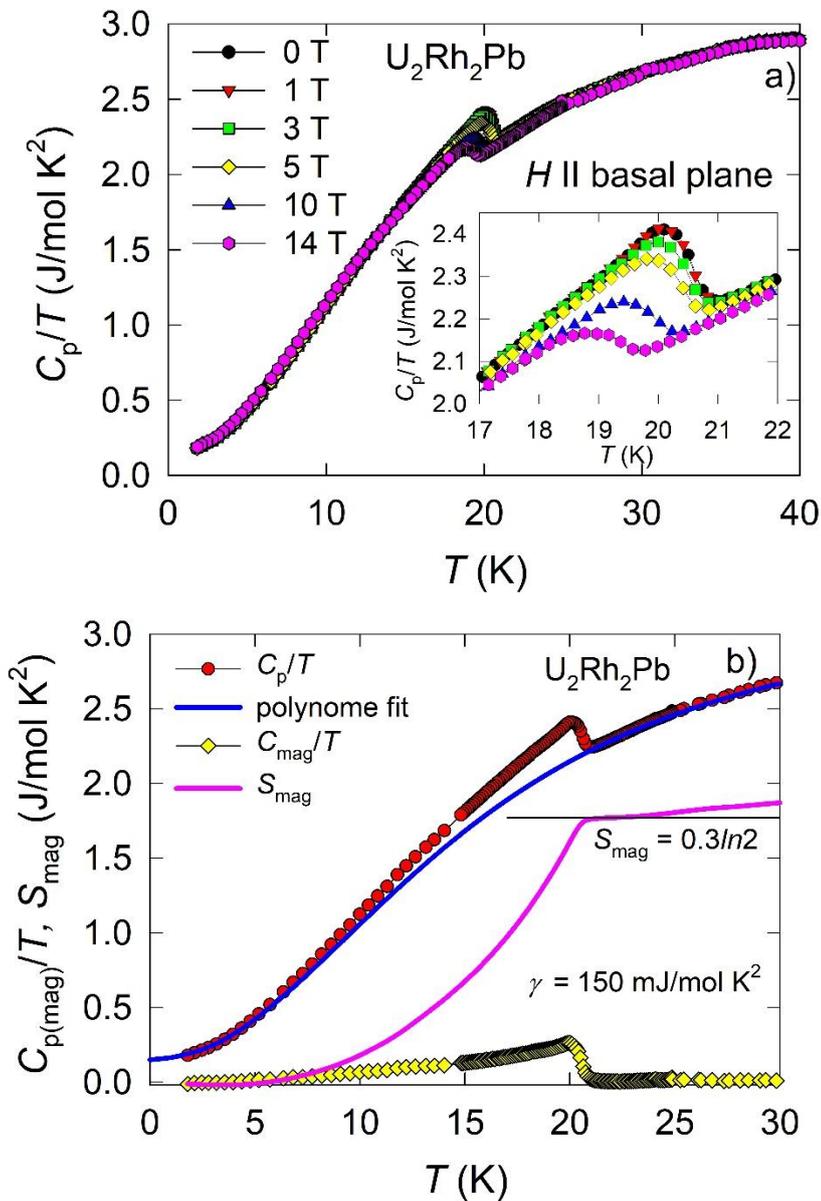

Fig. 2. a) Temperature dependence of the specific heat (plotted as $C_p/T$) of U$_2$Rh$_2$Pb single crystals in magnetic field parallel to the randomly oriented basal planes The inset shows the effect of the magnetic field on the value of $T_N$ in detail. b) Analysis of the zero-field specific heat with evaluated values for $\gamma$ and $S_{mag}$.

In contrast, if the magnetic field is applied along the $c$-axis, the anomaly shifts much faster to lower temperatures with increasing field, resulting in vanishing antiferromagnetism in magnetic fields exceeding 4.3 T (Fig. 3). In this range all specific-heat curves coincide. This is also the case for all curves below 10 K, with or without applied field, which indicates a negligible difference of the $\gamma$ coefficient values for the low-temperature magnetic states below and above the critical magnetic field $H_c$.

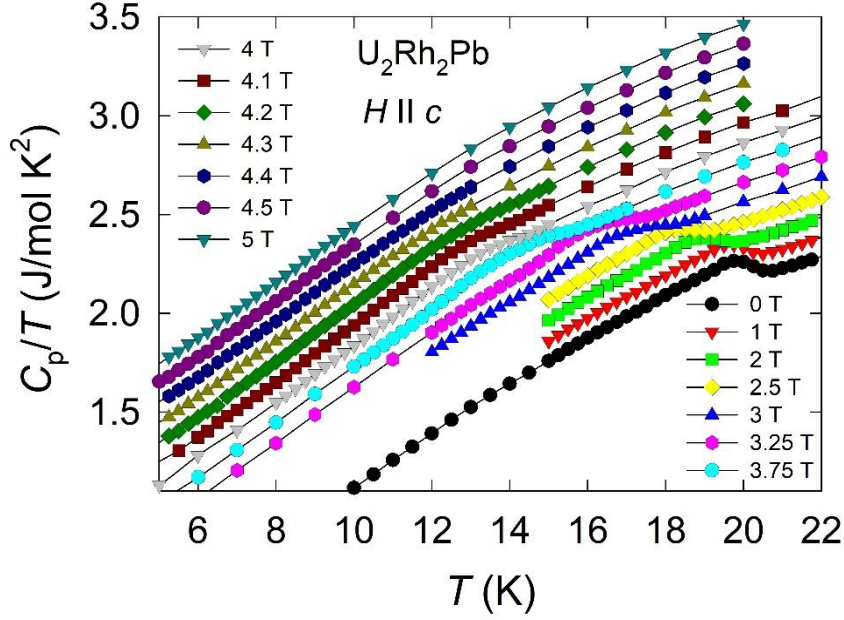

Fig. 3. Temperature dependence of the specific heat (plotted as $C_p/T$) of U$_2$Rh$_2$Pb in magnetic field applied along the $c$ axis. For clarity, the curves have been shifted along the $C_p/T$ axis by about 0.1 J/mol K$^2$ with increasing value of the applied field.

Specific-heat isotherms in applied magnetic field are of importance for construction of the magnetic phase diagram [30] and for estimation of transformations of the Fermi surface [31]. Surprisingly, it is found for U$_2$Rh$_2$Pb that, below 10 K, there is no sign of a clear anomaly and the isotherms are almost field independent (Fig. 4). The field-induced transformation from the AFM state to the polarized paramagnetic state is visible as a broad-peak anomaly. Above 16 K, the anomaly changes from peak-like to step-like and the state below $H_c$ has a higher value of $C_p/T$ than the state above $H_c$. While a clear step at $H_c = 1.5$ T is still observed at 20 K, no anomaly is present anymore at 21 K.

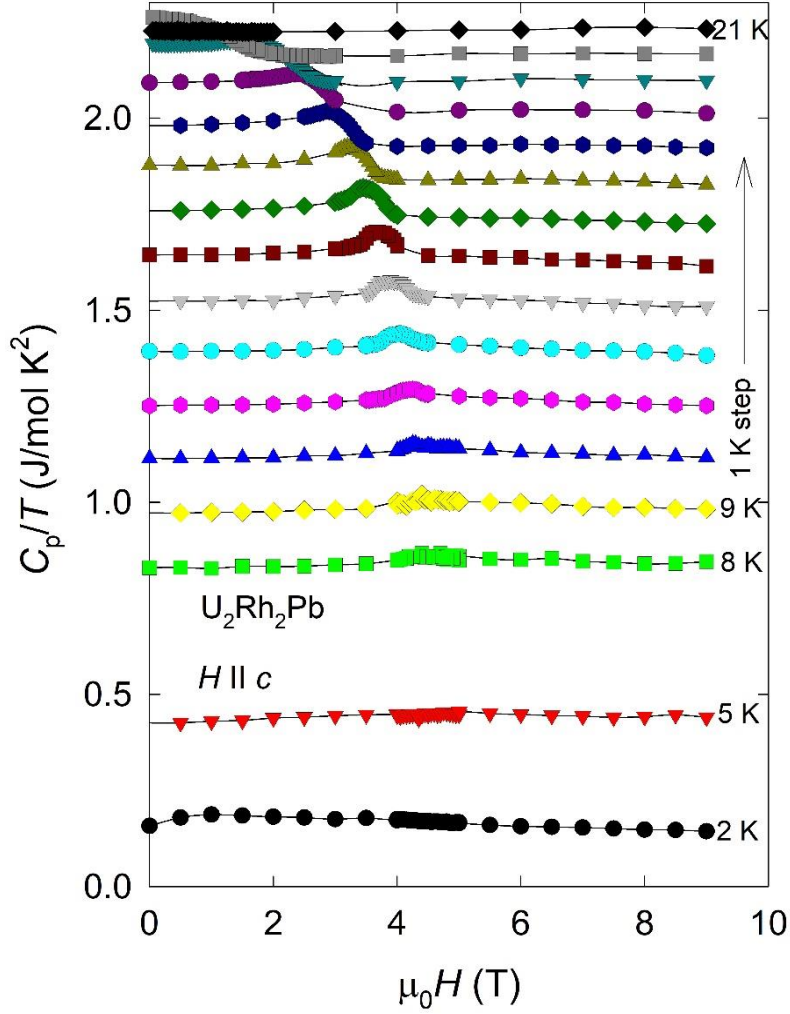

Fig. 4. Dependence of the specific heat (plotted as $C_p/T$) of $U_2Rh_2Pb$ on magnetic field applied along the $c$-axis at various temperatures below $T_N$ and at one temperature (upper curve) above $T_N$.

3.3 Magnetization

The magnetization measurement of very tiny needle-like crystals was rather difficult and systematic error up to 30 % of the value of the magnetization in Figs. 5 and 6 cannot be excluded. However, the magnetization signal was well detectable and a detailed magnetization study along the easy magnetization direction could be performed. The AFM order of $U_2Rh_2Pb$ indicated by the specific-heat results is confirmed by the temperature dependence of the magnetization (susceptibility) which displays a sharp maximum at $T_N = 20$ K (Fig. 5a) in magnetic field along the $c$-axis. We assign the weak increase of magnetization as well as the small anomaly at $T_N$ if the field is applied parallel the basal plane to a small misalignment of needles and projection of the easy axis. In the paramagnetic region, already from temperatures just above $T_N$, the temperature dependence of the reciprocal magnetic susceptibility $1/\chi$ is well described by the Curie-Weiss (CW) law giving $\theta_p = 1$ K, $\mu_{eff} = 2.6$ $\mu_B$/f.u., i.e. 1.84 $\mu_B$/U, which is significantly lower than $U^{3+}$ or $U^{4+}$ free-ion values [32] supporting the itinerant character of the 5$f$ electrons. The value of the magnetization in fields along the basal plane is very low and, due to the low mass of the very thin needle-shaped single crystals, the experimental detection limit was approached in the paramagnetic region. Therefore, CW analysis was not possible and we could not quantify the magnetic anisotropy in the paramagnetic state.

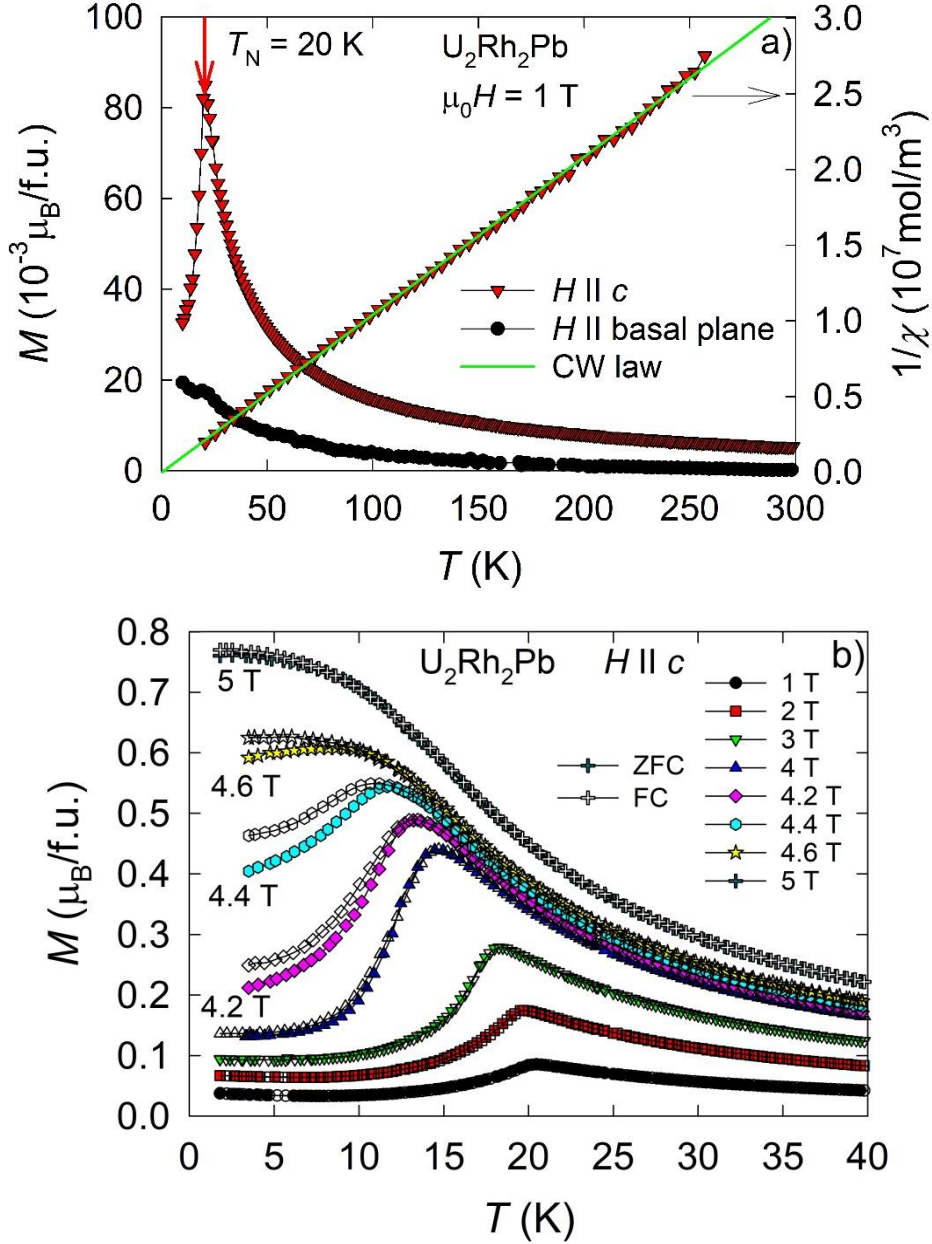

Fig. 5. a) Temperature dependence of the magnetization for magnetic field applied along the $c$-axis. The derived inverse susceptibility is fitted by the CW law. b) FC and ZFC magnetization curves measured at various magnetic-field values. The clear difference between FC (empty points) and ZFC (filled points) curves below $T_N$ appears in magnetic fields between 4 T and $H_c$.

Figure 5b shows the bifurcation between the field-cooled (FC) and zero-field-cooled (ZFC) branches occurring in magnetic fields above 4 T, which can be connected with emergent hysteresis at the MT. Another feature is a shallow minimum in the low-field magnetization curves around 7 K. We do not have a clear explanation but an impurity contribution due to residual remaining flux on the single crystals is the most likely scenario. Moreover, the minimum coincides with the observed drop in the electrical resistivity due to the superconducting transition of residual Pb nodules.

The magnetic-field dependence of the magnetization of $U_2Rh_2Pb$ at 1.8 K is presented in Fig. 6a. The low magnetization and absence of any sign of a jump of the magnetization observed for the field parallel to the randomly oriented basal planes indicate strong magnetocrystalline anisotropy, which is in agreement with the observations in the specific heat. The magnetization isotherms confirm that the *c*-axis is the easy-magnetization direction. In Fig. 6a can be seen that, at 1.8 K, a MT of the spin-flip type occurs at the critical field $\mu_0 H_c$ = 4.3 T applied parallel to the *c*-axis. The MT is accompanied by a weak hysteresis of about 25 mT which disappears at temperatures above about 6 K. The observed value for $\mu_0 H_c$ of 4.3 T is so far the lowest among the AFM $U_2T_2X$ compounds [2], for which the values of $H_c$ are typically larger than 20 T and field-induced transitions may appear even above 60 T [33].

Increasing the temperature up to 10 K has only a weak effect on the value of the critical field $H_c$ (Fig. 6b). Above 10 K, the MT becomes broad and $H_c$ shifts to lower fields. The metamagnetic jump disappears at $T_N$ above which the magnetic isotherms have a Brillouin character. It is interesting to compare the behaviour in fields of $U_2Rh_2Pb$ with isostructural rare-earth based compounds, as $Nd_2Ni_2Mg$ [34] or $Tb_2Pd_2Mg$ [35]. Although the MT occur already in fields below 10 T, their magnetic phase diagram contains multiple ordered phases extending to much higher fields. The only so far documented compound with easy reorientation of magnetic moments is $Nd_2Ni_2In$ with moments turned from the basal plane into the *c*-direction in a spin-flop transition completed in 0.3 T [36]. The uniaxial magnetic behaviour with low critical field and one magnetic phase in $U_2Rh_2Pb$ remains so far unique.

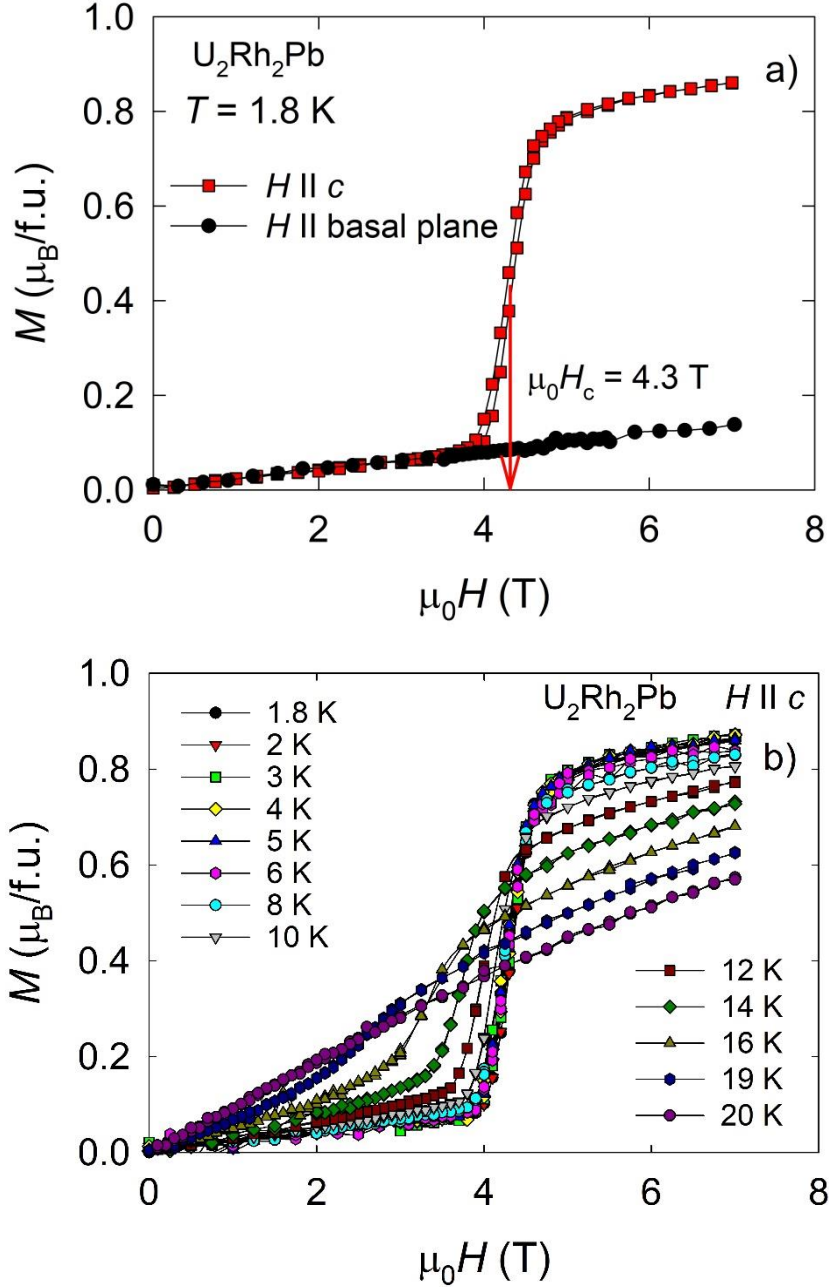

Fig. 6 a) Magnetization of $U_2Rh_2Pb$ measured at $T = 1.8$ K with the magnetic field parallel to the $c$-axis and parallel to the randomly oriented basal planes.

The $c$-axis orientation of U moments in $U_2Rh_2Pb$ is in contradiction with the tendency of uranium magnetic moments oriented due to the two-ion anisotropy perpendicular to the shortest $d_{U-U}$ distance [27, 37] which is along the $c$-axis in $U_2Rh_2Pb$. Following this rule, all the $U_2T_2X$ compounds with the shortest $d_{U-U}$ along $c$ have an easy-magnetization axis within the tetragonal basal plane [26, 38] and $U_2Ni_2Sn$ with the shortest $d_{U-U}$ within the plane has the moments along $c$ [37]. The only exception, $U_2Rh_2Sn$, which has also the easy magnetization direction along $c$, can be perhaps explained by a strong hybridization between U 5$f$ states and Rh 4$d$ electron states [27, 37]. In this context, $U_2Rh_2Pb$ with the c-axis U-U spacing 3% below the shortest U-U spacing in the basal plane is even more exceptional than $U_2Rh_2Sn$, where this difference is 2% at room temperature and 1% at low temperatures. However, the temperature variations of crystal structure of $U_2Rh_2Pb$ are still to be determined. The case of $U_2Ni_2Sn$ with *c* strongly increasing with decreasing *c* while *a* decreases in rather regular way shows that the lattice geometry can still undergo substantial changes below the room temperature.

3.4  Electrical resistivity

The needle-like morphology of the $U_2Rh_2Pb$ crystals is particularly suitable for electrical-resistivity measurements for the current in the *c*-direction. At zero field, the transition to the AFM state is detectable in the temperature dependence of the resistivity, $\rho(T)$, only as a feeble drop below $T_N$ (Fig. 7a). $\rho(T)$ was measured in various magnetic fields up to 5 T. The data largely overlap and $T_N$ is almost invisible (see inset of Fig. 7a). To establish $T_N$ more precisely, the minimum of $d\rho/dT$ was used (Fig. 7b).

In the paramagnetic state, $\rho(T)$ exhibits only a weak tendency to saturation, in a contrast with other compounds in the $U_2T_2X$ series, which exhibit a well-developed knee or even a negative slope at high temperatures. However, the absolute values at high temperatures exceeding 100 $\mu\Omega$cm are high and undoubtedly reflect presence of electron scattering on spin fluctuations. The residual resistance ratio RRR = 97 of the used needle significantly exceeds values of previously reported $U_2T_2X$ crystals, which likely points to a high perfection of the material prepared from flux. However, an influence of possible nodules of pure Pb, cannot be excluded. The influence of Pb superconductivity can be seen in the drop at ≈ 7 K, which can be suppressed by the field of 0.1 T to reveal an intrinsic behaviour.

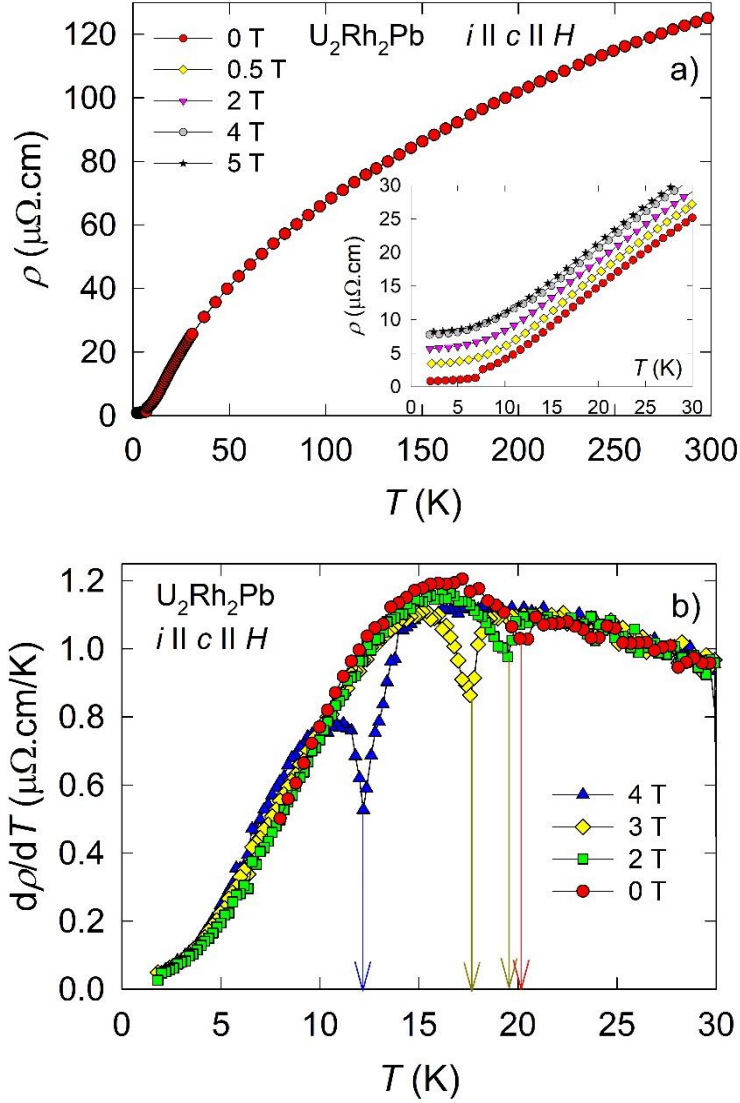

Fig. 7 a) Temperature dependence of the electrical resistivity in zero magnetic field. The inset shows some results of measurements below 30 K in various magnetic fields. For better clarity, the curves at various magnetic fields have been shifted by about 1 µΩ.cm. The drop of resistivity at 7.2 K in zero magnetic field is caused by the superconducting transition of spurious Pb. b) Derivative of the temperature dependence of the electrical resistivity. The arrows mark the values of $T_N$.

Interesting information can be in general obtained from $\rho(T)$ close to the low-$T$ limit, where electron-electron scattering term $AT^2$ can be observed in particularly in strongly correlated electron systems. However, we could not clearly detect the $\rho \sim T^2$ dependence, which suggests that low-lying magnetic excitations are present even below 10 K. A common approach is to use an additional exponential term describing magnon excitations over an energy gap, the energy of which can be associated with the magnetic anisotropy per ion.

Also an evaluation of the data with the Equation 1, that was used in the recently studied antiferromagnets $U_2Zn_{17}$ [39] and UIrGe [31] was not successful because unrealistic negative values for the parameter $A$ were obtained, even for various temperature intervals.

$$\rho = \rho_0 + AT^2 + BT\left(1 + \frac{2T}{\Delta}\right)exp\left(-\frac{\Delta}{T}\right) \quad (Eq.1)$$

Good agreement with reasonable values for the parameters was reached for the simple expression (Eq. 2)

$$\rho = \rho_0 + AT^2 + BT\exp\left(-\frac{\Delta}{T}\right) \text{ (Eq. 2)}$$

in which the last term represents the magnon contribution with the magnon gap $\Delta$.

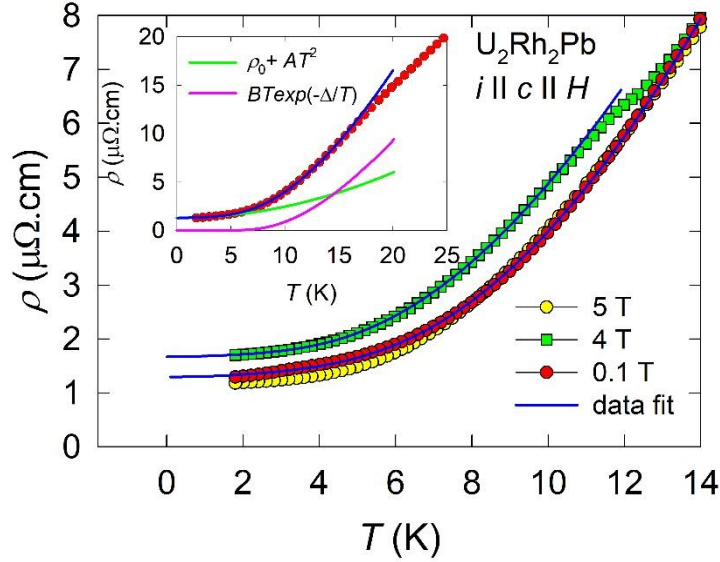

Fig. 8. Temperature dependence of the electrical-resistivity for some selected values of the magnetic field applied along the $c$-axis, together with fits to Eq.2. The inset shows separated contributions of each term in Eq. 2 represented in 0.1 T magnetic field data.

Some selected fitting results are presented in Fig. 8. It is important to note that the fits to Eq. 1 are the best at fields above 2 T. In the low-field data, some deviation can be seen, particularly at low temperatures (below 8 K). The fitting parameters are summarized in Table 4. Upon increasing magnetic field, $\rho_0$ increases but suddenly drops above $H_c$. The electron-electron correlations represented by the parameter $A$ remains almost constant. On the other hand, the parameters of the magnon term exhibit a clear minimum around $H_c$.

**Table 4**. Parameters obtained by fitting the electrical resistivity data to Eq. 1.

| $\mu_0 H$ [T] | $\rho_0$ [$\mu\Omega$.cm] | $A$ [$\mu\Omega$.cm/K$^2$] | $B$ [$\mu\Omega$.cm/K] | $\Delta$ [K] |
|---|---|---|---|---|
| 0.1 | 1.29 | 0.011 | 1.82 | 24.7 |
| 0.5 | 1.40 | 0.009 | 1.89 | 24.2 |
| 1 | 1.51 | 0.010 | 1.86 | 24.7 |
| 2 | 1.63 | 0.012 | 1.58 | 23.6 |
| 3 | 1.66 | 0.012 | 1.34 | 20.6 |
| 4 | 1.67 | 0.011 | 1.36 | 18.7 |
| 5 | 1.15 | 0.009 | 1.81 | 22.6 |

Verifying the validity of the empirical Kadowaki-Woods ratio [40] $A/\gamma^2 = 1.0 \times 10^{-5}$ $\mu\Omega$cm (molK$^2$/mJ)$^2$ for the new compound U$_2$Rh$_2$Pb with $\gamma \approx 150$ mJ/molK$^2$ and A $\approx 0.01$ $\mu\Omega$.cm/K$^2$, a significantly lower value of $A/\gamma^2 \approx 4 \times 10^{-7}$ $\mu\Omega$cm (molK$^2$/mJ)$^2$ is found. The deviation between the expected and evaluated ratio likely origin in complex Fermi surface structure of the anisotropic antiferromagnetic system represented by $U_2Rh_2Pb$ and parameter *A* can change for various electrical current directions.

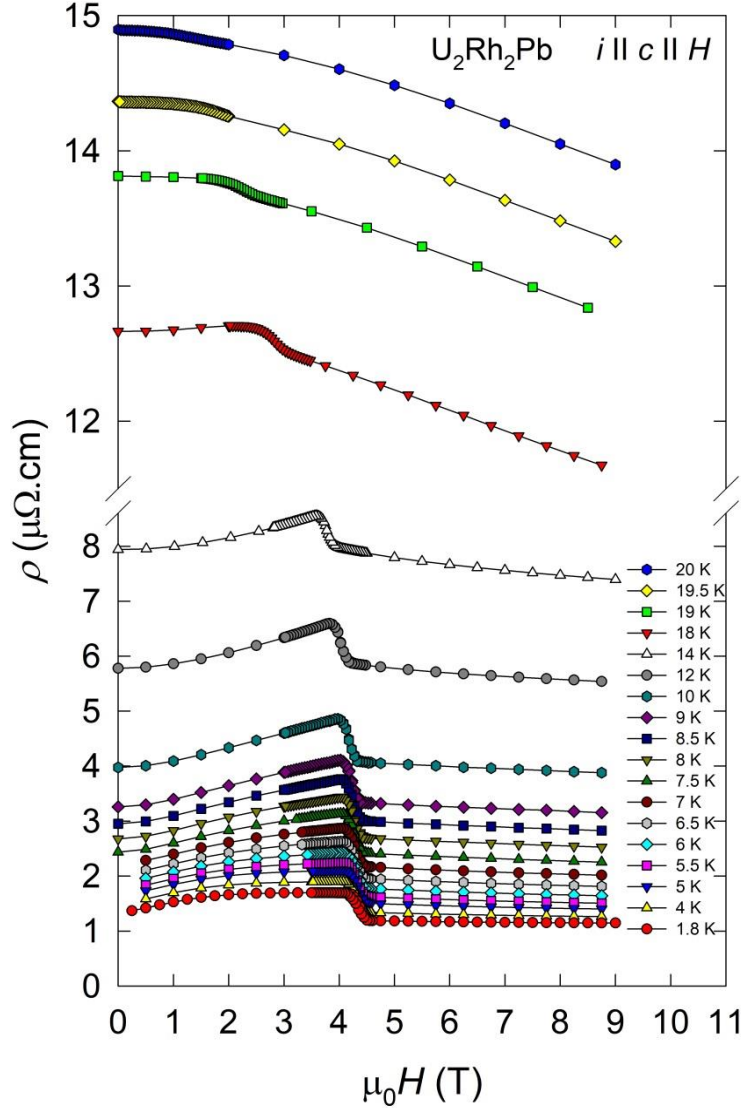

Fig. 9. Magnetoresistance of $U_2Rh_2Pb$ measured at various temperatures below $T_N$ and one above $T_N$. $H_c$ is well detectable as a drop of the electrical resistivity upon increasing field. Curves are arranged one by one in the order of the legend.

The magnetoresistivity in antiferromagnetic state exhibits a typical cusp at the metamagnetic transition, related to field-induced excitations in the sublattice with magnetization antiparallel to the applied field, which culminate just at the critical field. Only at very low temperatures, where the excitations are not assisted by thermal fluctuations, the anomaly converts to a resistivity step to lower values, typically related to the reconstruction of the Fermi level when the AF order vanishes. Close to $T_N$ and in the paramagnetic state the general negative magnetoresistance is due to the field effect on fluctuating moments existing also in the paramagnetic state.

Due to an abrupt decrease of the resistivity of $U_2Rh_2Pb$ at $H_c$, upon going from the AFM to the polarized paramagnetic phase state (Fig. 9), the magnetoresistance is a powerful tool for construction of the phase diagram of this compound. With increasing temperature, the value of $H_c$ shifts to lower fields. The effect of temperature is, however, very weak and, up to about 14 K, $H_c$ is still larger than 4 T. Approaching $T_N$, $H_c$ suddenly drops to zero and no trace of $H_c$ can be detected in the magnetoresistance curve at 20 K. A small hysteresis $\mu_0 H^{1.8K}_{hyst} \approx 50$ mT is detected at $H_c$ which is comparable with the observation in the magnetization. This hysteresis at $H_c$ disappears at temperatures above 13.2 K which is higher than for the magnetization data.

3.5 Magnetic phase diagram

The magnetic phase diagram of $U_2Rh_2Pb$ for $H$ along the $c$-direction is shown in Fig. 10. In contrast to many uranium antiferromagnets studied recently, the phase diagram of $U_2Rh_2Pb$ is rather simple. There is only one magnetically ordered AFM phase. A special feature is the occurrence of AFM transitions with and without hysteresis. The temperatures above which the hysteresis and the bifurcation of the FC and ZFC magnetization curves disappear are indicated in the phase diagram by arrows.

In previously studied AFM materials [41-43], the disappearance of the hysteresis has been considered as one of the signatures of a tricritical point (TCP) in the phase diagram which separates first- and second-order AFM transitions and exhibiting Ising-type anisotropy [42, 44-47]. This supports the existence of TCP in $U_2Rh_2Pb$ around $T_{tc} = 13$ K and $\mu_0 H_{tc} = 4$ T. It should be pointed out that TCP is an inherent feature of a broad class of antiferromagnets, in which the metamagnetic transition at low temperatures is of the first-order type and magnetic critical point in zero field is of the second-order type, and hence does now imply any temperature variations of magnetic interactions [48].

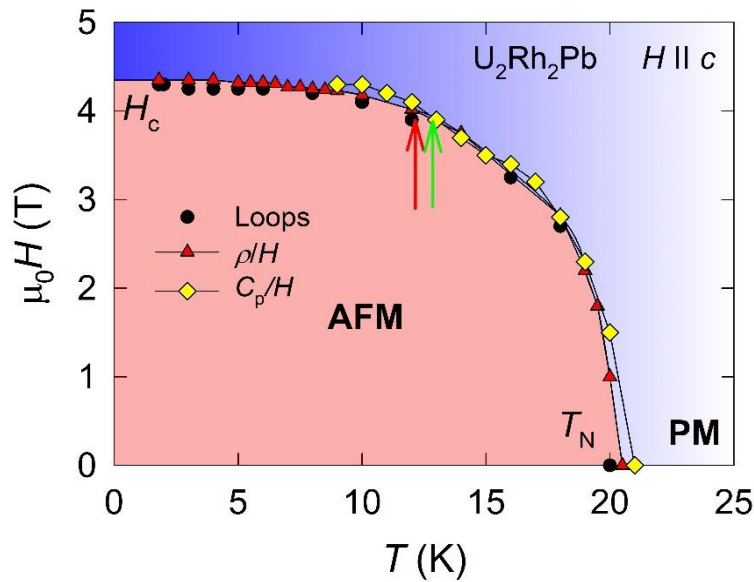

Fig. 10. Magnetic phase diagram of $U_2Rh_2Pb$ for $H$ parallel to the $c$-axis. The red arrow indicates the temperature above which the hysteresis in the magnetoresistance at $H_c$ disappears. The green arrow indicates the field above which clear bifurcation of the FC and ZFC magnetization branches is detected. The coloured crossover of the PM state tentatively marks polarization of the PM phase by external magnetic field.

3.6 Theoretical calculations based on DFT

Having no direct information on the type of magnetic structure, we used the Dirac-Kohn-Sham equations assuming a ferromagnetic arrangement. Magnetic ordering was indeed obtained and the values of the spin and orbital magnetic moments of $U_2Rh_2Pb$ were calculated-see Table 5 providing the value of 1.05 $\mu_B$ for the total moment per f.u.

**Table 5**. Calculated value of magnetic moment of $U_2Rh_2Pb$ decomposed to the spin $\mu_S$ and orbital $\mu_L$ magnetic moments for each element.

|  | $\mu_S$ ($\mu_B$) | $\mu_L$ ($\mu_B$) | $\mu_{total}$ ($\mu_B$) | $\mu_{total}$ $|(\mu_B/f.u.)|$ |
|---|---|---|---|---|
| U | 1.418 | -1.756 | -0.338 | 0.676 |
| Rh | -0.140 | -0.028 | -0.168 | 0.336 |
| Pb | -0.041 | ~ -0.000 | -0.040 | 0.040 |
| $U_2Rh_2Pb$ |  |  |  | 1.052 |

Mainly the U 5f-states contribute due to the large spin and antiparallel orbital magnetic moments. The induced spin and orbital magnetic moment on Rh rhodium are smaller than 0.14 $\mu_B$ and 0.03 $\mu_B$, respectively. Spin and orbital magnetic moments of lead are negligible, having a value less than 0.04 $\mu_B$. The total calculated moment 1.05 $\mu_B$ is in reasonable agreement with the value obtained from the magnetization above the MT (see Fig. 6). The total energies of the ferromagnetic and a simple AFM structure (up-down) were calculated using the APW+lo method including spin-orbit coupling. The total energies are comparable, slightly in favour of FM arrangement, which would be in contrast with the experimental findings. To estimate the magnetocrystalline anisotropy energy (MAE), the total energies with uranium magnetic moment along the *a*- and *c*-axis were calculated using the Dirac-Kohn-Sham FPLO code. In accordance with the experimental findings, the *c*-direction was found to be an easier magnetization direction than the *a*-direction. As also indicated by the experiments, a large MAE equal to 19.6 meV/f.u. (i.e. ≈ 115 K per U atom) was found.

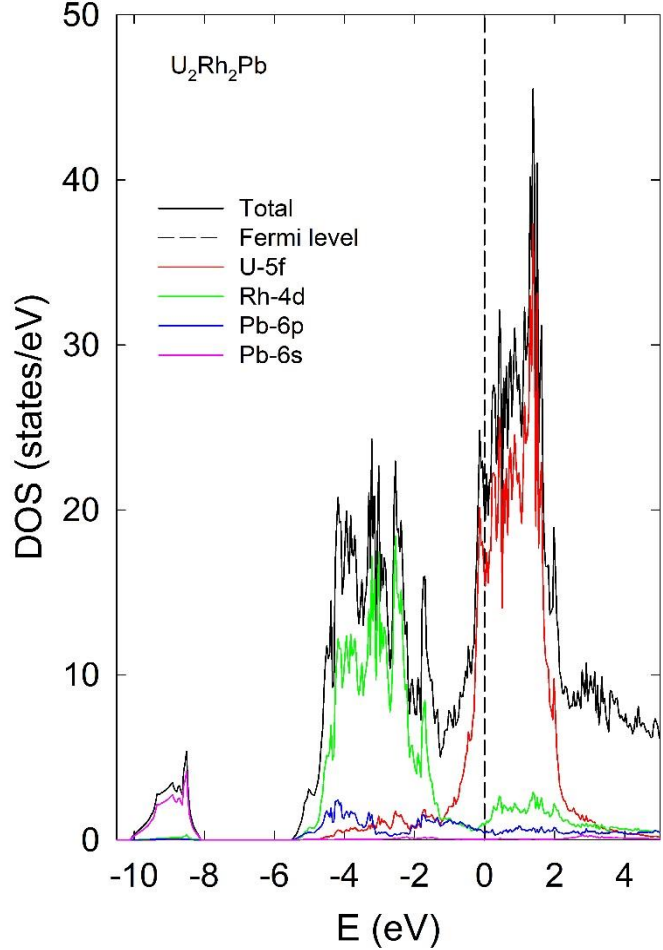

Fig. 11. Total and partial densities of states of $U_2Rh_2Pb$.

The fully-relativistic electronic structure (calculated by the FPLO code) of $U_2Rh_2Pb$ (see Fig. 11) displays a density of states (DOS) with the U 5f-band intersected by the Fermi energy $E_F$ and the Rh 4d-states are located around -3 eV, which is almost identical to the computed electronic structure of the heavy-fermion Rh analogue $U_2Rh_2In$, calculated using the Optimized Linear Combination of Atomic Orbitals (OLCAO) method [3].

It can be seen in Fig. 11 that there is strong hybridization between U 5f-states, Rh 4d-states and Pb 6p-states around -4eV and -2eV. The split-off of DOS from -10 to -8 eV corresponds to Pb 6s-states. The total DOS at the Fermi level equals 19.8 states/eV which corresponds to a calculated $\gamma \approx 23.3$ mJ/mol K$^2$. This is in strong disagreement with $\gamma \approx 150$ mJ/molK$^2$ derived from specific-heat data and reflects the limited validity of the single-particle description in the DFT framework. Electron-phonon interaction, many-body electron-electron correlations as well as pronounced magnetic-moment fluctuations are ill-treated in DFT but clearly present in $U_2Rh_2Pb$.

In order to compare the results of LSDA and GGA calculations, the variation of the total energy of $U_2Rh_2Pb$ with $V/V_0$ ($V_0$ is the experimental volume) has been calculated. The LSDA [20] value of the equilibrium volume is 5.8 % smaller than the experimental equilibrium volume which is the typical deviation obtained in LSDA full-potential calculations. GGA results for $V/V_0$ are found to deviate much less: 3.5 % smaller and 2.6 % smaller using the different GGA methods in Refs [23] and [22], respectively. The best results are obtained with the GGA method of Perdew et al. [21], which overestimates the experimental $V_0$ by only 1.4 %.

## 4  Conclusions

U$_2$Rh$_2$Pb, a new member of the large series of tetragonal U$_2$T$_2$X (T – transition metal, X – Sn, In, Pb) compounds, was synthesized in single-crystal form using the flux growth technique. It orders antiferromagnetically below $T_N$ = 20 K. The enhanced Sommerfeld coefficient $\gamma$ of about 150 mJ/mol K$^2$ infers the presence of strong correlation effects and the reduced magnetic-entropy change 0.3*$ln$2 at $T_N$ indicates itinerant character of 5f electrons. At low temperatures, the magnetization exhibits a spin-flip like MT at 4.3 T. The observed $c$-axis orientation of U magnetic moments is in contrast with the nearest U-U bonds along $c$, usually implying the basal-plane orientation. Magnetic phase diagram constructed for $H$ parallel to the $c$-axis hints to the existence of a tricritical point around 13 K and 4 T.

**Acknowledgments**

The research leading to the present results was conducted within the framework of the COST Action CA16218 (NANOCOHYBRI) of the European Cooperation in Science and Technology and the INTER-COST Project (No. LTC18024) of the Czech Ministry of Science and Education. The work of M. Diviš was supported by the Czech Science Foundation under Grant No. 18-02344S. The work of M. Dušek was supported by the Czech Science Foundation under Grant 18-10438S. The experiments were performed at MGML (http://mgml.eu), which is supported within the program of Czech Research Infrastructures (project no. LM2018096).